\documentclass[manuscript]{aastex63}
\usepackage{natbib}
\usepackage{graphicx}
\usepackage{soul}
\usepackage{xcolor}
\usepackage{amsfonts, amsmath, amsthm, amssymb}

\usepackage[super]{nth}
\providecommand{\e}[1]{\ensuremath{\times 10^{#1}}}

\usepackage{color}

\begin{document}

\title{A Practical Guide to the Partition Function of Atoms and Ions}

\author{P. Alimohamadi}
\author{G. J. Ferland}
\affil{University of Kentucky, Lexington, KY 40506, USA}

\begin{abstract}
The partition function, $U$, the number of available states in an
atom or molecules, is crucial for understanding the physical state of any astrophysical system in thermodynamic equilibrium.
There are surprisingly few {\em useful} discussions of the partition function's numerical value.
Textbooks often define $U$; some give tables of
representative values, while others do a deep dive into
the theory of a dense plasma.
Most say
that it depends on temperature, atomic structure, density,
and that it diverges, that is, it goes to infinity, at high temperatures, but few give
practical examples.
We aim to rectify this.
We show that there are two limits, $1$~\& $2$ electron (or closed-shell) systems like H or He,
and species with a complicated electronic structure like C, N, O, and Fe.
The high-temperature divergence does not occur for $1$~\& $2$ electron systems in practical situations since, at high temperatures, species are collisionally ionized to higher ionization stages and are not abundant.
The partition function is then close to the statistical weight of the ground state.
There is no such simplification for many-electron species.
$U$ is temperature-sensitive across the range of temperatures where an ion is abundant but remains finite at
even the highest practical temperatures.
The actual value depends on highly uncertain truncation theories in high-density plasmas.
We show that there are various theories for continuum lowering but that they
are not in good agreement.
This remains a long-standing unsolved problem.

\end{abstract}

\clearpage

\tableofcontents
\thispagestyle{empty}
\clearpage

\section{Introduction}
\label{Introduction}
\setcounter{page}{1}

The partition function $U$ is the number of available states of an atom or molecules.
It is at the heart of
computing the Local Thermodynamic Equilibrium (LTE) level occupation of atoms and ions using the Boltzmann equation and the ionization of each chemical element using the Saha equation.
Calculating its numerical values is not a simple task because of
the well-known divergence at high temperatures, which is discussed below.
Also, despite its critical importance,
we know of no general introduction to the numerical value of
the $U$ for atoms and ions
(we do not consider molecules here).
This tutorial aims to this.

Here, we provide a brief survey of discussions of the
partition function in astrophysical textbooks. 
\citet{1996ima..book.....C} gives an equation for the partition function, but with no more detail.
\citet{blundell2010concepts} defines the partition function as the sum over the Boltzmann factor without considering the degeneracy of each level determined by the statistical weight. This is a hypothetical two-level system introduced only for tutorial purposes. 
\citet{2014aspr.book.....B} lists the partition function as just due to the ground state. 
At the opposite extreme, \citet{2014tsa..book.....H} gives an extensive summary of the theory described by \citet{1988ApJ...331..794H}, but it provides no numerical values,
while \citet{2005pps..book.....G} presents an even more formal discussion from the laboratory plasma perspective.
Some books 
\citep{1968asa..book.....S, 1973itsa.book.....N, 1973asqu.book.....A, 2000asqu.book.....C} give tables of values for selected species over specific temperatures, but there is no detail of the procedures used nor guidance for the density or temperature dependence.
\citet{1973itsa.book.....N} and \citet{1968asa..book.....S}, 
give $U$ and comment that the reported values are valid 
regardless of temperature if the ion is abundant.
\citet{1968AcSpe..23..521D, 1984ApJS...56..193S} give polynomial expressions for the partition function versus temperature, which give approximate estimates of $U$, but do not describe
the density dependence. 

our goal is to discuss the densities and temperature dependencies
of atomic and ionic partition functions.
We begin by considering the simplest and most important case,
atomic hydrogen. 
We demonstrate its divergence over a very larger temperature range and demonstrate
its divergence.
We give an overview of some theories
for how to treat the very high states that cause the function
to diverge.
The Saha equation shows that atomic hydrogen will not be abundant
when the temperature is high enough for $U$ to diverge.
This simplifies the partition function for most densities and temperatures since it becomes the statistical weight of the ground state.
We then go on to consider many-electron systems with their
complex energy structure.  
A robust and reliable theory of continuum-lowering and dense-plasma effects
is needed to handle finite densities properly.
These effects limit the number of states that contribute to $U$ and prevent its divergence.
There are different physical theories, used in different communities,
which do not agree well.

\section{The partition function in astrophysics}
\label{sec: Partition functions in astrophysics}

\subsection{Notation in this tutorial}
\label{subsec: Notation in this paper}

The current convention in astrophysics is to express column density 
as $N$ [cm$^{-2}$] and the particle number-density as $n$ [cm$^{-3}$].
Here we use $n$ for the principal quantum number and $N$ for the number-density. 

There are different notations, such as $Q$, $Z$, and $U$ for the partition function. We use $U$ for the partition function.  

\subsection{Range of density and temperature}
\label{subsec: Range of density and temperature}

We discuss temperatures ranging between $2.7~\mathrm{K}$ and $10^8~\mathrm{K}$.
The temperature $T = 2.7$~K is the lowest temperature encountered in the cosmos, the cosmic microwave background temperature, a near-perfect black-body.
The upper limit is so high that the most elements
will be fully ionized so the properties of bound atomic states
are not a concern.

Our density range extends from $N_{\mathrm{e}}= 10^{0}~\mathrm{cm^{-3}}$, essentially the low-density limit,
up to
$10^{25}~\mathrm{cm^{-3}}$, a density well above
the lower regions of the atmospheres of accretion disks
or stars.
The partition function would normally be used in systems close to
LTE, which requires higher densities
\citep{2014tsa..book.....H}. 
Low-density gas is so far from equilibrium as to make many LTE concepts irrelevant \citep{2006agna.book.....O}.
\newpage
\subsection{Definition of the partition function}
\label{subsec: Definition of the partition function}

The partition function is 
the sum over all quantum states of an atom
with a degeneracy or statistical weight of~$g_n$:
\begin{equation}
\label{eq: Partition function} 
U = \sum_{n=1}^{\infty}g_n\exp\Big(\frac{-E_n}{k_{\mathrm{B}}~T}\Big)
\end{equation}
\noindent where ${\exp\;(-E_n/k_{\mathrm{B}}~T)}$, the Boltzmann factor, gives the probability of the presence of an electron in some particular state, $n$, 
at temperature $T$, which has the energy of $E_n$. 
For a hydrogenic system, the statistical weight is
\begin{equation}
\label{eq: Statistical weight}
g_n = 2 n^2
\end{equation}
The energy of the $n$-th shell for a hydrogenic atom,
governed by the static screened Coulomb potential approximation,
is
\begin{equation}
\label{eq: Electronic Energy}
E_n = Z^2 I_\textrm{H}\;\Big(1 - \frac{1}{n^2}\Big)
\end{equation}
\noindent where $I_{\mathrm{H}}$ is the ionization energy of the hydrogen atom, $I_{\mathrm{H}}=~2.17\;\e{-11}$\;erg.

\begin{figure}
\centering
\includegraphics[scale=0.9]{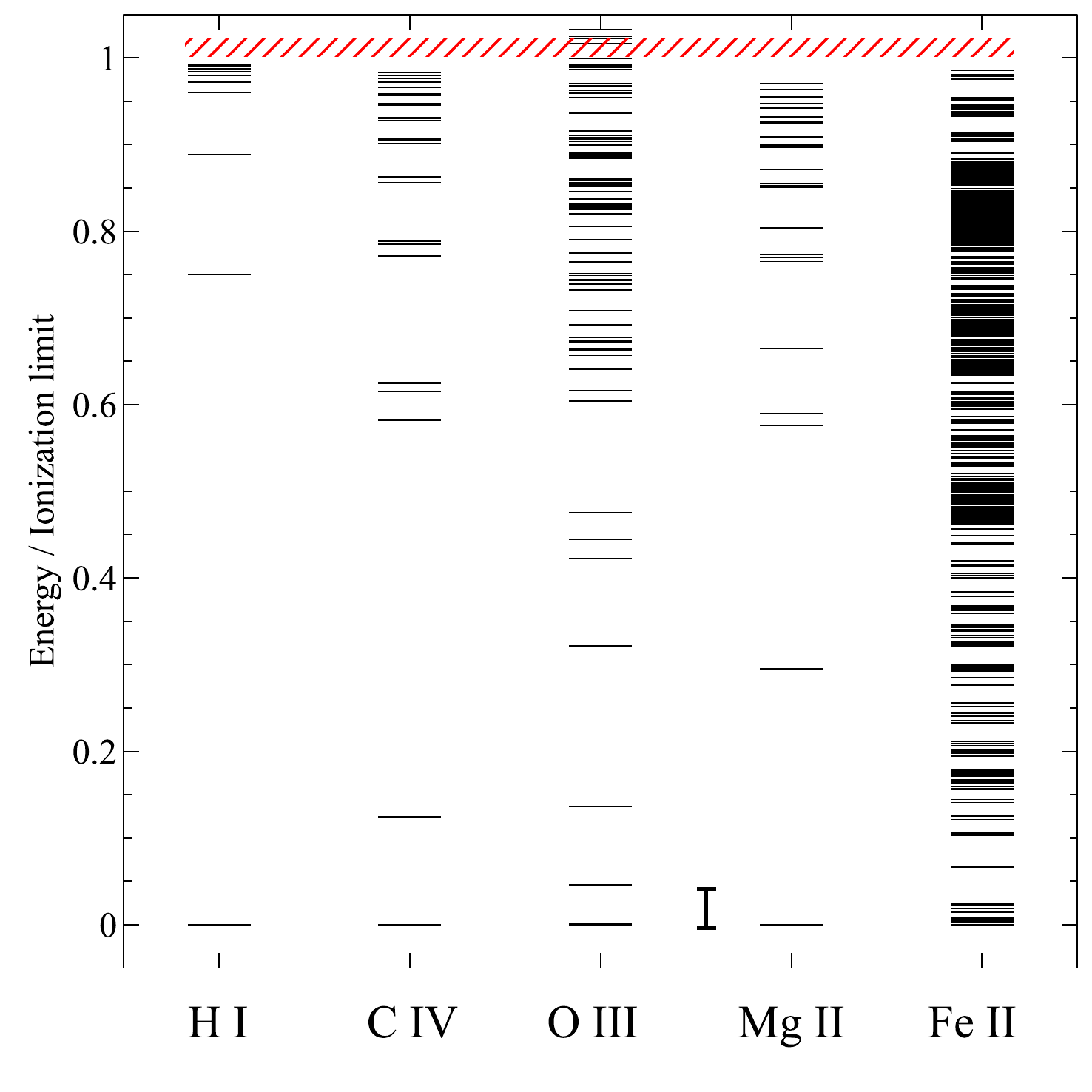} 
\caption{Experimental energy levels relative to each elements' ionization potential for some species presented in \citep{2014APS..DMP.D1047K} and adapted from \citet{2017RMxAA..53..385F}. 
The red hashed lines show the scaled ionization limit. 
Those lines above the ionization limit shown for O~III are auto-ionizing levels. 
As it is shown, for hydrogenic ions, most 
\nth{1} and \nth{2} excited states are close to the continuum 
compared to the many-electron ions, 
where they are close to the ground state.
The vertical line between O III and Mg II indicates an energy
corresponding to $\sim 10^4$~K.}
\label{figs: EnergyLevels}
\end{figure}

For reference throughout this tutorial, Figure~\ref{figs: EnergyLevels} shows energy levels 
of different species with an increasing number of orbiting electrons
going from left to right.  
The level energies are given relative to the ionization energy,
the red band at $y=1$.  
Simple systems like H~I have energy levels given by Equation \ref{eq: Electronic Energy}, and have a simple structure.
Complex systems have a far higher density of lower-energy states due to the complex interactions
between various orbiting electrons.
There is an infinite number of levels for all systems since $n$ can, in theory, extend to infinity.

\subsection{Ionization and excitation - the Boltzmann and Saha equations}
\label{subsubsec: the Boltzmann and Saha equations}

The partition function plays a pivotal role in computing
the LTE ionization and level excitation using 
the Saha and Boltzmann equations. 
The LTE ionization of atoms is given by the Saha equation, 
which, for the case of the hydrogen atom, is
\begin{equation}
\label{eq: Saha}
\frac{N(\mathrm{H~II})}{N(\mathrm{H~I})} =  \frac{2}{N_{\mathrm{e}}}\Big(\frac{2 \pi m_{\mathrm{e}} k_{\mathrm{B}} T}{h^2}\Big)^\frac{3}{2} \;\frac{U(\mathrm{H~II})}{U(\mathrm{H~I})}\;\exp \Big(\frac{-I_{\mathrm{H}}}{k_{\mathrm{B}} T}\Big)
\end{equation}
\noindent where $U(\mathrm{H~I})$ and $U(\mathrm{H~II})$ represent the partition function for
H~I and H~II. 
The Boltzmann constant, electron mass, Planck constant, and temperature are represented by 
$k_{\mathrm{B}}$, $m_{\mathrm{e}}$, $h$, and $T$, respectively. $N_{\mathrm{e}}$ is the electron density [$\mathrm{cm^{-3}}$]. 
Considering only atoms and ions, the conservation equation becomes
\begin{equation}
\label{eq: charge conservation}
N(\mathrm{H}) = N(\mathrm{H~I}) + N(\mathrm{H~II})
\end{equation}
\noindent allowing us to compute the ionization fractions of H~I and H~II relative to the total hydrogen density, $N({\mathrm{H})}$.   

The Boltzmann equation gives the LTE level populations of atoms in 
a given excited state at some temperature.
The population of the  $n$-th shell
will be
\begin{equation}
\label{eq: Boltzmann eqaution}
\frac{N_{n}(\mathrm{H~I})}{N(\mathrm{H~I})} = 
\frac{g_{n}}{U(\mathrm{H~I})}~
\exp \Big(\frac{-E_{n}}{k_{\mathrm{B}} T}\Big)
\end{equation}
where $E_{\mathrm{n}}$ is obtained by Equation \ref{eq: Electronic Energy}.
As shown in Equations \ref{eq: Saha} and \ref{eq: Boltzmann eqaution}, 
the partition function 
is required for the ionization and excitation calculations.
The partition function of H~II, a bare proton, 
is $U(\mathrm{H~II}) = 1$, a proton's statistical weight. 

\subsection{The partition function for an infinite and finite atom}
\label{subsec: The partition function for an infinite and finite atom}

This section outlines the behavior of the partition function. It is organized as follows:

The upper limit of the summation in Equation~\ref{eq: Partition function}
includes all possible states.
First, we show that this partition function diverges,
it goes to infinity, for all temperatures where there is an infinite number of states.

We then numerically evaluate the partition function across a wide 
temperature range for a large but finite number of levels. 
The partition function does not go to infinity 
but does become very large at high temperatures. 
At low temperatures it takes the value of the statistical weight of the ground state.
The very large-$n$ levels that cause the divergence have a nonphysically large radius, and cannot exist in practical circumstances.
This introduces the concept of the truncation of the partition function, the physics setting the highest-$n$ that actually occurs. This should be the upper limit of the summation in Equation~\ref{eq: Partition function}.
This limit to the highest principal quantum number is discussed in
Section~\ref{sec: An overview on the truncation of the partition function}.  

\subsubsection{The divergent partition function for an infinite level atom}
\label{subsubsec: The divergent partition function for an infinite level atom}

We first show that the partition function will diverge for any $T$ if the number of levels goes to infinity.
For large $n$, the excitation energy of H~I given by Equation~\ref{eq: Electronic Energy} becomes the ionization energy:
\begin{equation}
\label{eq: E for infty n}
\lim_{n \to \infty} E_n = \lim_{n \to \infty} I_{\mathrm{H}}\Big(1 - \frac{1}{n^2}\Big) \approx I_{\mathrm{H}}.
\end{equation}
At high-$n$ and $T$, where the kinetic energy $k_{B}~T$ is much greater than the ionization energy  $I_{\mathrm{H}}$, we have $k_{\mathrm{B}}~T \gg I_{\mathrm{H}} 
\Longrightarrow k_{\mathrm{B}}~T \gg E_n$.
The Boltzmann factors are all nearly unity, $\exp{(I_{\mathrm{H}}/k_{\mathrm{B}} T)} \rightarrow 1$, since the argument in the exponential goes to zero.
Then, $U_n \approx 2n^2$, the statistical weight of the $n$ configuration,
and $U=\sum_1^\infty U_n \rightarrow \infty$. 

At low temperatures the term in the exponent in 
Equation~\ref{eq: Partition function} is 
large and negative.
The Boltzmann factor becomes small, though finite. 
Therefore, no matter the temperature, the terms contributing to the partition function given by Equation~\ref{eq: Partition function} become infinity
\begin{equation}
\label{eq: analytic U}
\lim_{n \to \infty} U_n \approx
\lim_{n \to \infty} 2n^2 \exp{(-I_{\mathrm{H}}/k_{\mathrm{B}}~T)} 
\Longrightarrow \infty .
\end{equation}
The partition function diverges for all temperatures if the
sum over principal quantum number is extended to infinity.

\subsubsection{The partition function for a finite atom}
\label{subsubsec: The partition function for a finite atom}

This section evaluates $U$ for a large but finite number of levels.
It shows that $U$ does not go to infinity; instead, it takes large values dominated by the highest levels.
This introduces the concept of the truncation of the partition function.
Here, we consider the partition function of H~I at low-density and a broad range of $T$ for large but finite numbers of levels to evaluate $U$.

\begin{figure}
\centering
\includegraphics[scale=0.8]{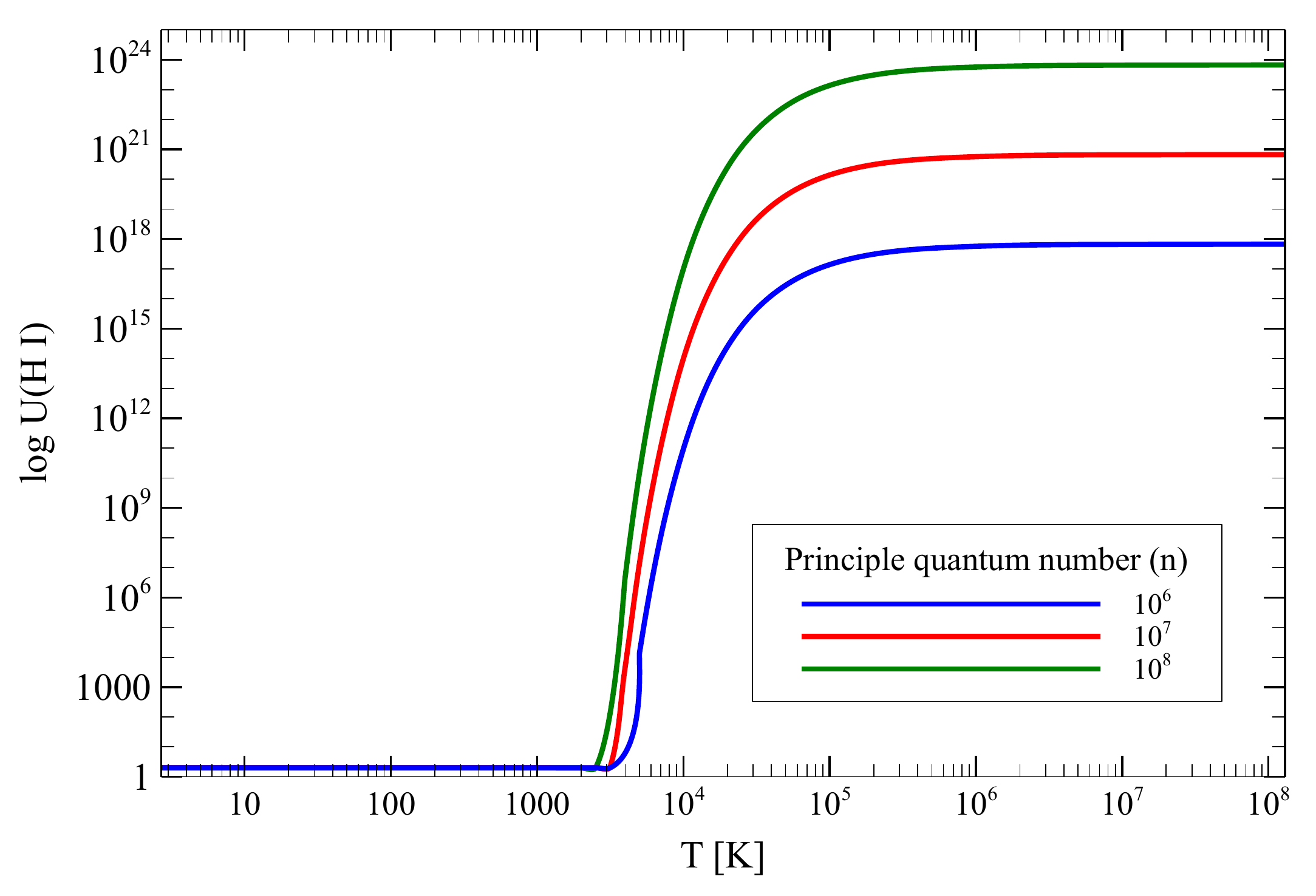} 
\caption{Partition Function versus limit to number of shells $n$ and temperature. 
This shows $U(\mathrm{H~I})$ versus $T$ for H~I
with the sum
in Equation~\ref{eq: Partition function} 
extending to $n=10^6$, $10^7$,~and~$10^8$. 
At low-$T$, the partition function is the statistical weight of the ground state, $g_n=2$.
$U$ begins to increase when the temperature is high enough to populate
the $n=2$ configuration.
At high temperatures, $U$ 
is dominated by the highest $n$, which 
theoretically could go to infinity, demonstrating its divergent behavior.}
\label{figs: Divergent PF}
\end{figure}

Figure~\ref{figs: Divergent PF} shows a numerical evaluation of $U(\mathrm{H~I})$ versus $T$
for three limits to the highest-$n$ to the summation in Equation~\ref{eq: Partition function}.
The partition function remains finite for all $T$.
At low-$T$, the Boltzmann equation (given by Equation~\ref{eq: Boltzmann eqaution}) suggests that nearly all the atom's population is in the ground state. So only $n=1$ contributes to the sum in Equation~\ref{eq: Partition function}.
Consequently, the partition function goes to $U(\mathrm{H~I})\approx 2$, the statistical weight of the ground state.  

The Boltzmann factors are all near unity at high $T$, so $U$ becomes the sum of $2n^2$. This converges onto a high but finite plateau. 
The Boltzmann equation suggests that the probability of populating excited states increases as temperature increases, and so physically, nearly all the atom's population is in the highest levels.
Therefore, the sum of $2n^2$ is dominated by the largest $n$, which accounts for the value of the plateau.

The infinite-$n$ model in the previous section helps explain
some of the behaviors in Figure~\ref{figs: Divergent PF}.
The Figure shows that
the lowest temperature where $U$ increases above $2$
depends on the limit to the number of levels. 
So, we can estimate that extending $n \rightarrow \infty$ will increase the height of the plateau at high-$T$ and move the values in low-$T$ to the left, and so $U(\mathrm{H~I})$ will go to infinity for the entire range of $T$. 

Very large-$n$ energy levels do not occur in nature
because these orbits would have a nonphysically large radius.
At low temperature, atoms mostly will be in the ground state, and an electron would have the most probable distance from the nucleus of one Bohr radius,
\begin{equation} 
\label{eq: Bohr radius}
a_0 = \frac{4\pi \epsilon_0 \hbar^2}{m_{\textrm{e}}~e^2} = \frac{\hbar}{m_{\textrm{e}} c \alpha} \approx 5.29~\e{-11}~[\textrm{m}] = 5.29~\e{-9}~[\textrm{cm}] = 0.529~[\textrm{\AA}]
\end{equation}
\noindent where $e$ is the elementary charge, and $\alpha$ is the fine structure constant, a dimensionless quantity independent of the system of units, and approximately is $\frac{1}{137}$. 

At high temperatures, $U$ is dominated by the highest $n$.  
The radius of an atom in the $n^{\textrm {th}}$ shell of a hydrogenic ion with charge $Z$ is
\begin{equation}
\label{eq: radius-n}
r = a_0 \Big(\frac{n^2}{Z}\Big)\approx 
5.29 \e{-11} \Big(\frac{n^2}{Z}\Big) [\textrm{m}] = 
5.29 \e{-9} \Big(\frac{n^2}{Z}\Big) [\textrm{cm}] = 
0.529 \Big(\frac{n^2}{Z}\Big) [\textrm{\AA}].
\end{equation}
For a hydrogen atom with level numbers of $n = 10^6$, $n = 10^7$, and $n = 10^8$ that we discussed in Figure~\ref{figs: Divergent PF}, Equation \ref{eq: radius-n} gives radii of
$r\approx53$~[m], $r\approx53\e{2}$~[m], and $r\approx53\e{4}$~[m] respectively.
These radii are far larger than  the separations between hydrogen atoms at realistic densities.  
These orbits do not exist so $U$ is finite. 

Setting the upper limit to the sum giving $n$ is the problem of
establishing its truncation.
The ``hard sphere'' model is the simplest approach
for truncating the partition function.
Here, the largest orbits are those that ``fit'' in the space between atoms.
The following sections outline this, and more sophisticated, prescriptions
for the truncation.

\section{The truncation of the partition function}
\label{sec: An overview on the truncation of the partition function}

The previous section reviewed the cause of the partition function's divergence at higher temperatures. Current approaches for its truncation, as introduced in plasma or chemical references,
are described here.

\subsection{Inter-particle interactions with Sharp cut-off}
\label{Inter-particle interactions with Sharp cut-off}

In Section~\ref{subsubsec: The partition function for a finite atom}, we showed that for high temperatures, there are a large fraction of electrons in large orbits, which causes $U$ given by Equation \ref{eq: Partition function} to diverge.
The physical problem is that
the very large orbits cannot exist due to interactions with nearby atoms in a plasma.
Taking these inter-particle interactions into account,
one would set an upper limit for the number of states of an atom. 
The ``hard sphere'' model of the truncation is one example
of how to do this.  It
can be visualized as the largest possible orbit that
can fit between neutral and charged particles in a plasma.

The revised
definition of the partition function is \citet{2002AcA....52..195H}
\begin{equation}
\label{eq: Revised partition function_sharp_cutoff} 
U = \sum_{n=1}^{\mathrm{n_{max}}}\;g_n\exp{\Big(\frac{-E_n}{k_{\mathrm{B}} T}\Big)}
\end{equation}
In other words, interactions with neighboring particles in the plasma decrease the ionization potential $I_{\mathrm{ion}}$ by some amount of $\Delta \chi$, and so, the maximum energy of the largest bound orbit is
\begin{equation}
\label{eq: max-energy}
E_{\mathrm{max}} = I_{\mathrm{ion}} - \Delta \chi
\end{equation}
where $\Delta \chi$ represents the lowering of the ionization energy,
called ``continuum lowering''.
The goal is to find a theory that gives $\Delta \chi$ as a function
of density or temperature. 
This type of theory gives a sharp cut-off to the divergence of the partition function.

There has been a tremendous effort to find a theory
that gives a correct form of $\Delta \chi$ as a function of density and temperature.
Some reviews of continuum-lowering theories due to Debye shielding, Stark, and collisional broadening
have been developed especially within the plasma physics community and
are described in
\citet{1939ApJ....90..439I, 1948ZA.....24..355U, 1978stat.book.....M, 1997PhLA..231...82H, 1998Ap&SS.259..173M, 2000ApJ...544..581B, 2005pps..book.....G, 2012fapc.book.....C}
or \citet{2020arXiv201110603K}.
These papers touch many different fields, ranging from laboratory plasmas to stars and astronomical plasma simulation codes like XSTAR \citep{2000ApJ...544..581B} or Cloudy \citep{2017RMxAA..53..385F}.

One example of these theories,
Equation~$9-106$  in \citet{1978stat.book.....M},
gives a continuum-lowering criterion due to Debye shielding effect.
The continuum is lowered by
\begin{equation}
\label{eq: delta_x Mihalas}
\Delta \chi = 3~\e{-8}~Z~N_{\mathrm{e}}^{\frac{1}{2}}~T^{-\frac{1}{2}}~ \mathrm{[eV]},
\end{equation}
a function of temperature, $T$, ionic charge, Z, and electron density, $N_{\mathrm{e}}$.

We consider the case of H~I ($Z = 1$) with sufficient lowering
to decrease the ionization potential by $\Delta \chi = 1$~eV,
to include only the first $4$ configurations,
where $\mathrm{n_{max}}=4$ is obtained using Equation~\ref{eq: Electronic Energy}. 
At $T = 2\e{4}$~K, 
Equation~\ref{eq: delta_x Mihalas} gives the density of $N_{\mathrm{e}} = 2.2 \e{19}~\mathrm{cm^{-3}}$. 

Different theories predict that $\Delta \chi$ 
would have different density- and temperature- dependencies
and results. 
Equations of~$3$,~$4$,~and~$5$ of \citet{2000ApJ...544..581B} and
again in \citet{2020arXiv201110603K},
(from \citet{1939ApJ....90..439I, 1997PhLA..231...82H}) 
compare the continuum-lowering criteria for particle packing, 
Debye shielding, and Stark broadening.
For $\Delta \chi = 1$~eV,   
particle packing finds a density of
$N_{\mathrm{e}}=1.64\e{21}~\mathrm{cm^{-3}}$,
Debye shielding gives the density of
$N_{\mathrm{e}}=3.57\e{31}~\mathrm{cm^{-3}}$,
and the Stark broadening gives the density of
$N_{\mathrm{e}}=5.54\e{21}~\mathrm{cm^{-3}}$.
\emph{These differ by 10 dex}.
\citet{1998Ap&SS.259..173M} presents the continuum lowering criterion with the nearest neighbor approximation (from \citet{1948ZA.....24..355U}), which gives a density of
$N_{\mathrm{e}}=2.91\e{18}~\mathrm{cm^{-3}}$, \emph{three dex below the lowest of the above}.
The Debye shielding theory in \citet{2012fapc.book.....C},  
their equation~$8.4$, gives a density of $N_{\mathrm{e}}=1.32\e{20}~\mathrm{cm^{-3}}$.
Including Debye shielding, the theories scatter over $14$ dex in density. Excluding Debye shielding, the remaining theories scatter over $\sim 3.5$~dex. The reason for these 
divergent predictions
is the difficulty in accurately describing the many-body interactions in
a dense gas with a simple physical theory.

\subsection{Inter-particle interactions with smooth cut-off}
\label{Inter-particle interactions with smooth cut-off}

The inter-particle interactions discussed above 
sets an upper limit to the radius of an atom,
solving the divergence of the partition function.
However, the solution is not complete because interactions 
with neighboring ions perturb the electrons in lower orbits such that they become less bound to the nucleus.
The probability that an orbit is filled so fully 
contributes to the partition function decreases as the distance from the nucleus increases.
This revises the definition of the partition function to read \citet{1988ApJ...331..794H}
\begin{equation}
\label{eq: Revised partition function} 
U = \sum_{n=1}^{\mathrm{n_{max}}}W_n\;g_n\exp{\Big(\frac{-E_n}{k_{\mathrm{B}} T}\Big)}
\end{equation}
The occupation probability, $W_n$, is the probability that an electron
can be in the $n$-th orbit due to density effects.
The numerical values of $W_n$ start from $1$ 
for fully occupied unperturbed levels and gradually decrease 
for the higher energy states and 
finally it goes to zero, $\lim_{n\to\infty} W_n = 0$
for large orbits that cannot exist. 
Some references, notably \citet{1960BAN....15...55D}, refer to $W_n$ as the perturbation function while 
\citet{1988ApJ...331..794H},
the most highly cited paper in this field, calls it the occupation probability. 

A detailed study of the inter-particle interactions is required to 
determine the occupation probability. 
There are two types of interactions between particles 
within a plasma. 
The first determines the largest orbit for an atom by taking the separation between atoms into account and assuming the other interacting particles are in the ground state.
The largest radius is related to the separation between particles.
This is known as the \emph{hard-sphere model} where the interactions are defined between neutral particles and is describe in
\citet{2014tsa..book.....H}
and \citet{1988ApJ...331..794H}. 
A numerical illustration of the  hard-sphere model 
is presented in figure~$8.14$ of \citet{2012fapc.book.....C}.  

The second model 
considers the electric field of these particles.
That perturbs the higher atomic energy levels and leads to their depopulation. 
To know the details of these perturbations, one needs to obtain the electric field distribution of the charged particles in a plasma. 
The Holtsmark distribution function is a convenient way to describe interactions between all charged particles \citep{1986JQSRT..36....1H}.
However, it does not consider the motion of the particles, which means that it neglects the perturbations due to the magnetic field of the charged particles.
As shown recently by \citet{2020A&A...635A.180V}, the magnetic field of the charged perturbers disturb the higher energy levels as well.

Two theories describe $W_n$ based on the inter-particle interactions,
the \emph{perturbation function model} \citep{1960BAN....15...55D}, 
and the \emph{occupation probability formalism} \citep{1988ApJ...331..794H}. 
The perturbation function model was developed to explain the gradual
dissolution of the Lyman lines of the hydrogen atom as an analogy to the gradual depopulation of the higher energy levels due to their perturbations by ionized or neutral particles. 
Here, higher-$n$ lines become indistinct and eventually disappear with the highest
$n$ depending on the density of the star's atmosphere. 
For more details see  \citet{1960BAN....15...55D}.

The occupation probability model \citep{1988ApJ...331..794H} describes the dissolution of the higher energy levels with the Stark effect and is by far the most widely used approach 
in stellar atmospheres.
The occupation probability function takes the form
\begin{equation}
\label{eq:Occupation Probability}
W_n = \int_{0}^{\beta_n} d \beta\;\;P_{\mathrm{\tilde{H}}}(\beta)
\end{equation}
where $P_{\mathrm{\tilde{H}}}(\beta)$ is the Holtsmark function 
\citep{1986JQSRT..36....1H} and $\beta_n$ is the reduced field strength. The final form of the occupation probability for both neutral and charged perturbers is given by Equation~4.71 of \citet{1988ApJ...331..794H}. 
Their equation~$4.42$ gives the maximum level-number 
for a hydrogen-like system, with $W_{\mathrm{n_{max}}} = {\mathrm{e}}^{-1}$ and $\beta = 1.8$, by  
\begin{equation}
\label{eq: maximum n} 
\mathrm{n_{max}} = 1.2 \e{3}~N_{\mathrm{e}}^{-2/15}~Z_{\mathrm{a}}^{3/5}
\end{equation}
where $Z_{\mathrm{a}}$ is the ionic charge of the perturbers, and $N_{\mathrm{e}}$ is the number density of electrons. 
For more detail, we refer to \citet{1988ApJ...331..794H}.

The next section considers $U$ for H~I but
takes into account
the ionization determined by the Saha equation.
The ionization establishes an upper limit to the temperature 
where an ion has a significant abundance, providing another
solution to the divergence of $U$ at high temperatures.

\section{One and two-electron systems; a simple limit}
\label{One and two-electron systems and a simple limit}

We show next that the electronic structure of hydrogenic systems,
together with the Saha equation,
provide a simple solution to the divergence of $U$ - namely,
the partition function of $1$~\&~$2$ electron systems is well-behaved for temperatures where an ion has a large abundance. 

\subsection{Asymptotic model with the Saha equation}
\label{subsec: Asymptotic Model with Saha equation}

\begin{figure}
\centering
\includegraphics[scale=0.4]{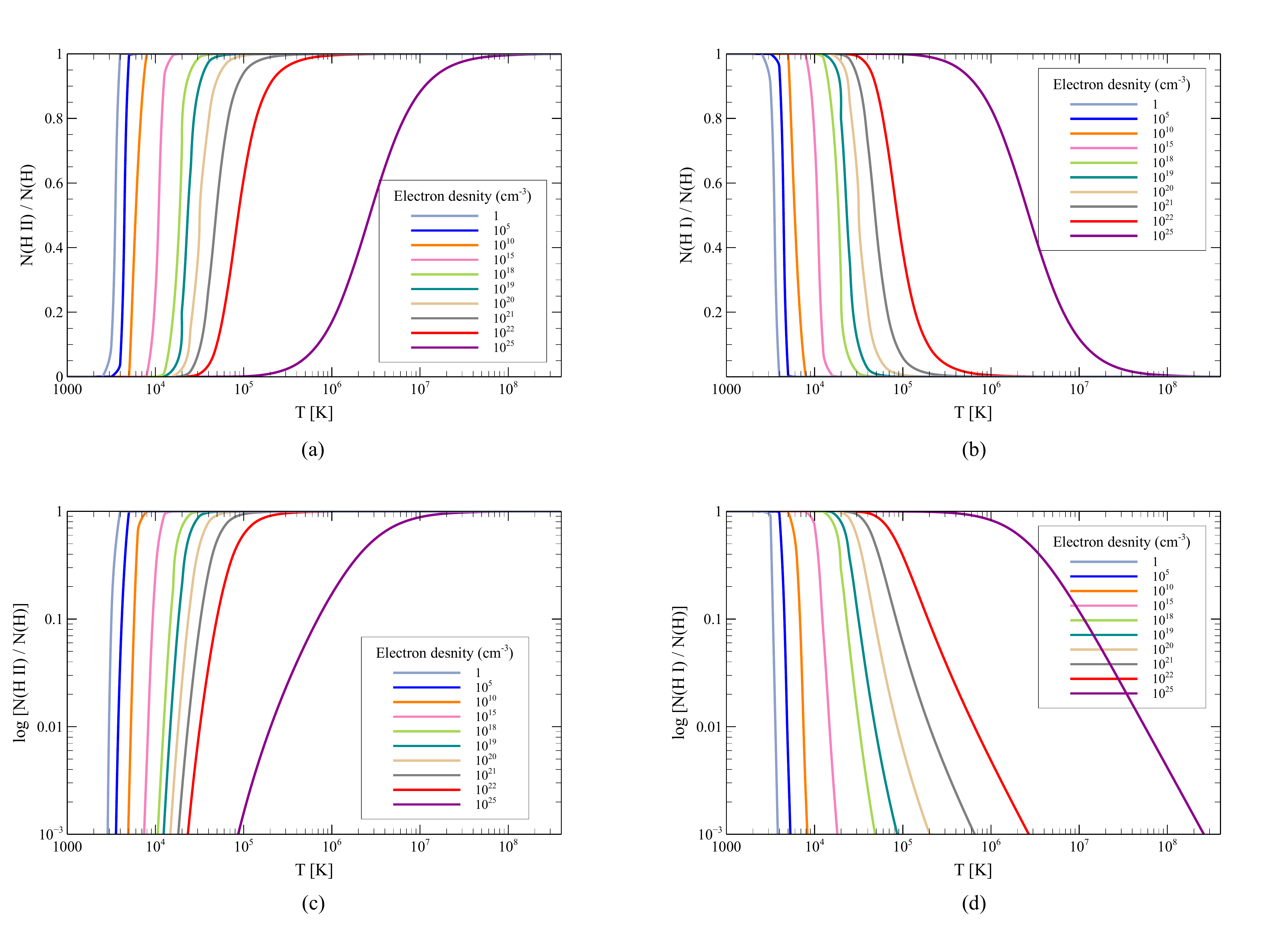} 
\caption{Saha Ionization Balance. 
($\mathrm{a}$): H~II fraction in linear scale, 
($\mathrm{b}$): H~I fraction in linear scale, 
($\mathrm{c}$): H~II fraction in logarithmic scale, ($\mathrm{d}$): H~I fraction in logarithmic scale. }
\label{figs: Saha}
\end{figure}

We use Equation~\ref{eq: Revised partition function} 
to compute $U(\mathrm{H~I})$ with the upper limit of the summation given by Equation~\ref{eq: maximum n} where $(Z_{\mathrm{a}} = 1)$. 
Since $W_n = 1$ is used for all energy levels, this is sometimes called
the \emph{“asymptotic version"} of the revised partition function. 
Figure \ref{figs: Saha} shows the LTE ionization fraction for 
H~I and H~II for a broad range of densities and temperatures.
Figures~\ref{figs: Saha}a~and~\ref{figs: Saha}c in linear and logarithmic scales respectively give results for the \emph{ionic} ionization fractions. Figures~\ref{figs: Saha}b~and~\ref{figs: Saha}d in linear and logarithmic scales respectively give the \emph{atomic} ionization fractions.

These figures establishes the temperature range where 
H~I is abundant, and $U(\mathrm{H~I})$ matters. 
The highest practical temperature is the value where a significant amount of H~I
is still present.
To set an upper limit to the
temperature range, we use H~I $/$ H $\leq 0.001$,
based on our Saha equation calculations. This limit could extend to H~I~$/$~H~$\leq~10^{-5}$ without any significant changes in the established temperature ranges.  
For these calculations, numerical values
of $U(\mathrm{H~II}) = 1$ and $U(\mathrm{H~I}) = 2$
were used in Equation~\ref{eq: Saha}. 

Figure~\ref{figs: Saha} demonstrates that for lower electron densities,
the hydrogen atom becomes ionized at quite low temperatures. 
As the electron density increases,
the temperature needed to fully ionize hydrogen rises. 
The lower limit to the y-axis of Figures~\ref{figs: Saha}c and \ref{figs: Saha}d is set to
$10^{-3}$ to focus attention on the temperature range where H~I is abundant. 
At these temperatures, H~II becomes abundant, which is shown in Figures~\ref{figs: Saha}a and \ref{figs: Saha}c.   
We also compute the correct H~I partition function using Equation \ref{eq: maximum n} as represented with the thin lines in Figure~\ref{figs: PF}, which we will
discuss later to derive an improved Saha ionization distribution.
There is only a slight deviation between these Saha calculations for the range of electron densities from $10^{18}~\mathrm{cm^{-3}}$ to $10^{21}~\mathrm{cm^{-3}}$ 
due to using the uncertain truncation criterion for computing the partition function by Equation \ref{eq: maximum n} or setting $W_n=1$. 
However, for most densities and temperatures, this does not change our results.

In the next section, we 
evaluate $U(\mathrm{H~I})$ versus temperature limits 
obtained by the Saha equation and show
that $U(\mathrm{H~I})$ does not diverge for most practical circumstances. 
\newpage
\subsection{A practical upper limit to the temperature}
\label{subsec: Results}

\begin{figure}
\centering
\includegraphics[scale=0.63]{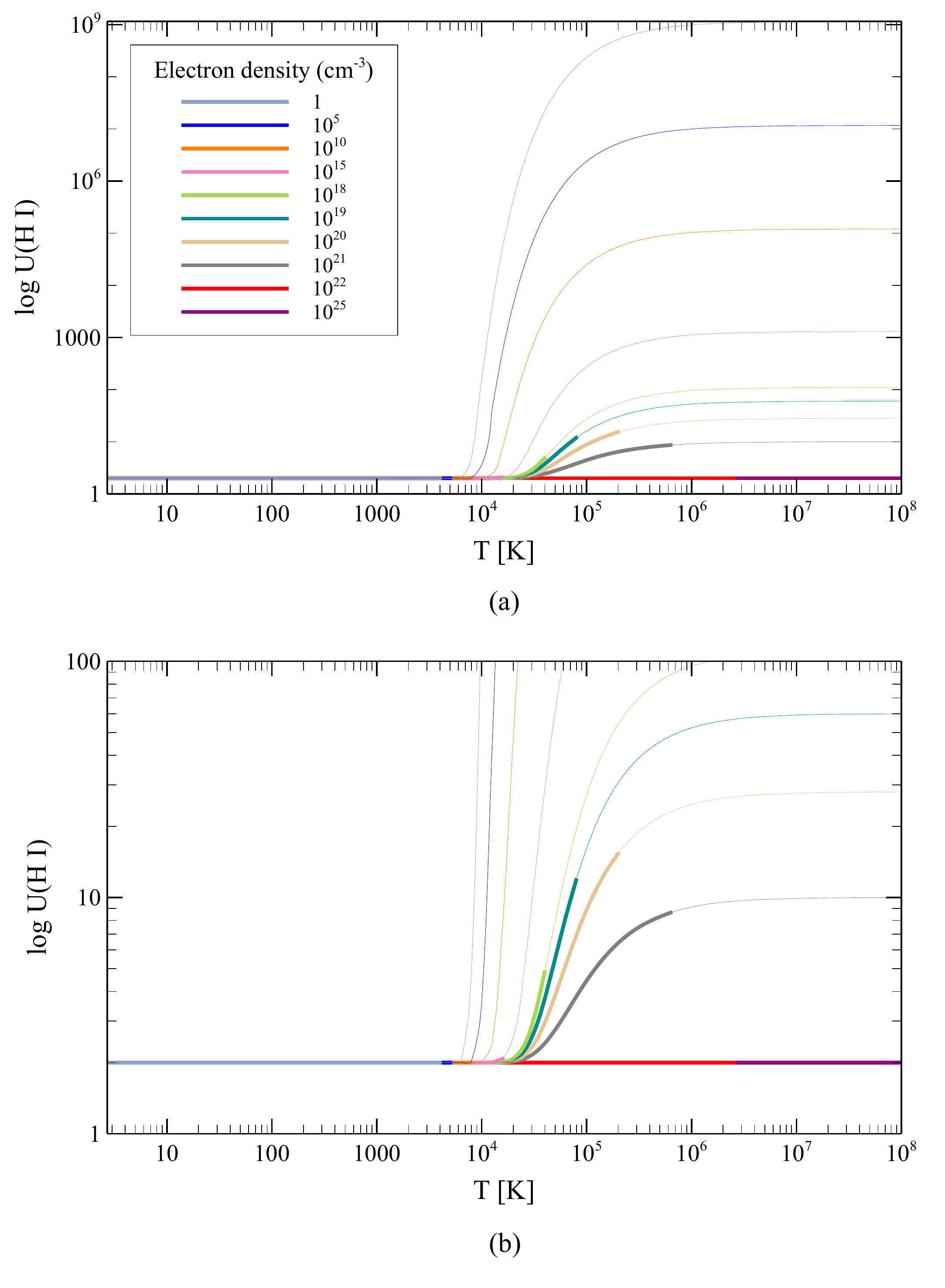} 
\caption{The partition function of atomic hydrogen. 
The thin lines represent $U(\mathrm{H~I})$ for the
full temperature range.
The thick lines indicate
the temperature range 
where H~I $/$ H $\geq 10^{-3}$. 
H is ionized to form H~II at higher temperatures. 
Although the eye is drawn to the thin lines, for most
temperatures and densities $U\sim 2$ and only rarely
does it exceed 10.}
\label{figs: PF}
\end{figure}

Figure~\ref{figs: PF}
shows $U$ over the same range of temperatures and electron densities considered
in Figure~\ref{figs: Saha}. 

The thin lines in Figure \ref{figs: PF} show $U(\mathrm{H~I})$ across for the full range. 
These lines deviate from the ground state statistical weight,~$2$, at moderate temperatures for the lower electron densities.
The thick lines show $U$ over the temperature range where H I is abundant.
The temperature limits established 
in Section \ref{subsec: Asymptotic Model with Saha equation} 
were used to assess this. 
Figure~\ref{figs: PF} shows that for most densities and temperatures
where H~I is abundant, so the results are given by thick lines, $U \approx 2$. 
Only for a narrow range of densities,
$10^{18}~\mathrm{cm^{-3}} - 10^{21}~\mathrm{cm^{-3}}$, is there a range of temperature where $U$ is larger than 2.
But, even in this range, $U$ is not very large, $U(\mathrm{H~I}) \leq 10$.

Figure~\ref{figs: PF}b expands the vertical axis to make this clearer.
For very high densities
like $10^{22}~\mathrm{cm^{-3}}$ or larger, $U(\mathrm{H~I})$ does not deviate from the ground state statistical weight
for the entire range of temperatures
because of the extreme lowering of the highest levels given by Equation~\ref{eq: maximum n}.

We provide another interpretation for $U(\mathrm{H~I})$ 
in Figure~\ref{figs: PF} using the energy levels of H~I presented in Figure~\ref{figs: EnergyLevels}.
Figure~\ref{figs: EnergyLevels} shows that the lowest excited states of H~I are closer to the continuum than the ground state.
This means an electron in the third excited state, $n=4$,
has an energy $E_n \approx (1 - \frac{1}{n^2}) \approx 0.95$
of its ionization limit.
At low temperatures, it does not contribute to $U$ since its Boltzmann factor is small.  When the gas is hot enough to populate
$n=4$, it is also hot enough to collisionally ionize the atom.

In general, the energy structure of hydrogenic systems 
means that 
most excited states are closer to the continuum than 
to the ground state,  
so if the population of the highly excited states is likely,
then collisional ionization is too and the abundance is small.
When the abundance of the atom is large,
the highly excited states make a negligible contribution to the partition function. 

Our results for $1$~\&~$2$ 
electron systems 
will tend to hold for any species with a closed-shell 
electronic structure due to the similar energy structure.
For example, the 
one-electron systems: H~I, He~II, and Li~III with $1s$ configuration, 
two-electron systems: He~I, Li~II, Be~III, B~IV, and C~V with $1s^2$~configuration, and species with
one electron in the L or M shell: 
Ne~I, Na~II, and Mg~III with $1\mathrm{s}^2 2\mathrm{s}^2 2\mathrm{p}^6$ configuration 
are closed-shell and roughly behave as quasi-hydrogenic species with a similar explanation for the truncation of their partition function.  
Hence, the ground state approximation is an appropriate 
approximation for
the partition function of the hydrogenic systems, or any system with a similar energy structure like closed-shell elements, and so $U(\mathrm{H~I}) \approx 2$ for most practical conditions. 

This result explains the statements in some older texts
\citep{1973itsa.book.....N, 1968asa..book.....S} attributed to unpublished calculations of the Arthur. N. Cox, the author of \citet{2000asqu.book.....C}.
They note that the partition function of these species does not depend on temperature and density if the ions are abundant.

\section{Complex ions - here be dragons}
\label{sec: Complex ions - here be dragons}
We have shown that the truncation of the partition function is
is uncertain.
Considerable simplifications can be made in hydrogenic cases, although
$U(\mathrm{H~I})$, does increase for low densities and high temperatures. 

The situation is not so simple for many-electron systems.
The energy levels associated with these ions are more complex, as shown in Figure~\ref{figs: EnergyLevels}.
Fe~II is a well-studied system that is observed in many astronomical
sources. 
It has a complex energy structure, and
here we consider it in detail.

\begin{figure}
\centering
\includegraphics[scale=0.8]{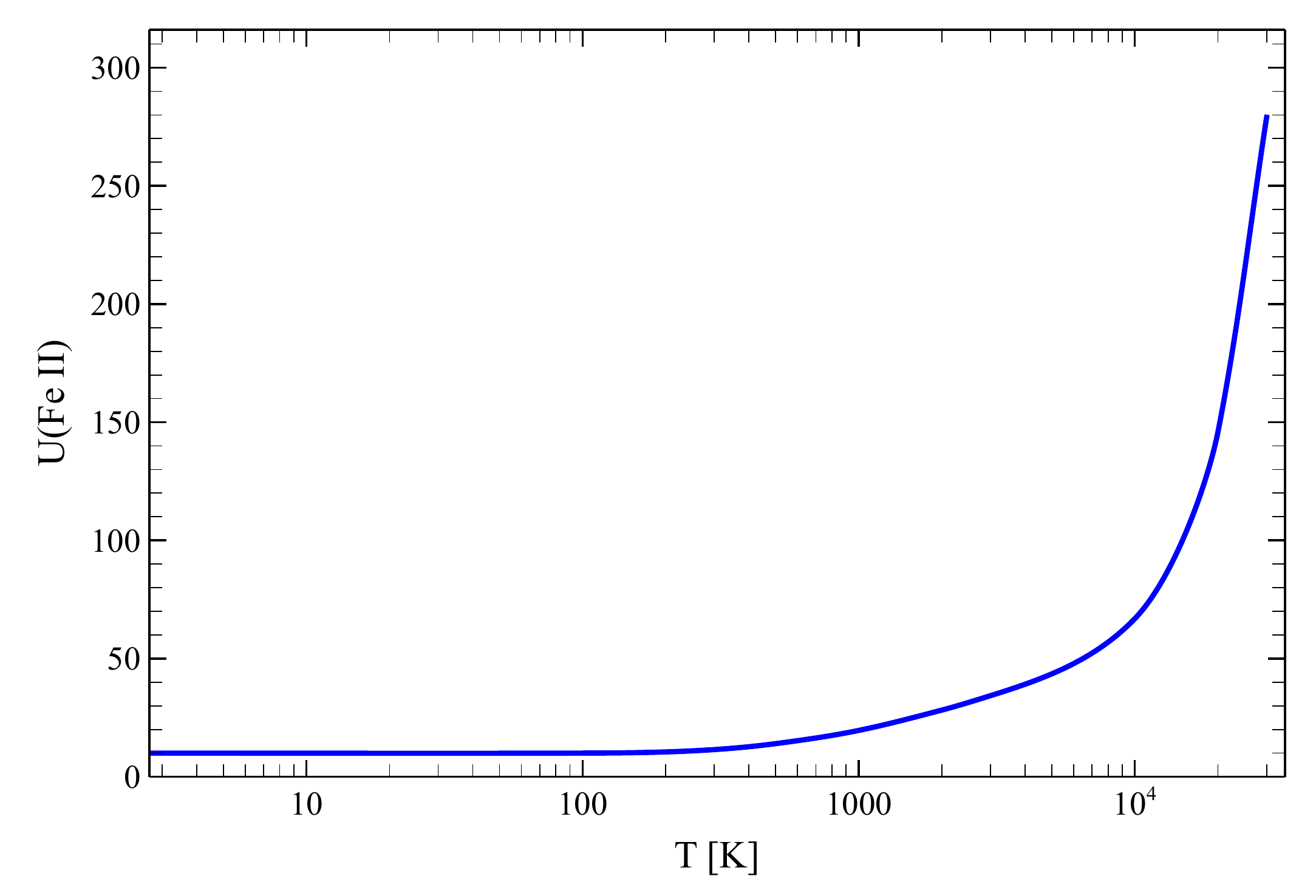} 
\caption
{The partition function of Fe~II. 
$U(\mathrm{Fe~II})$ is evaluated, 
including all energy levels presented in \citet{2019MNRAS.483..654S}, 
over the temperatures where the Fe~II abundance 
is significant. 
The high-temperature limit is where
Fe~II $/$ Fe~$\leq 0.01$ since Fe~II is collisionaly ionized
to form Fe~III.} 
\label{figs: U(FeII)}
\end{figure}

Figure~\ref{figs: U(FeII)} shows $U(\mathrm{Fe~II})$
as a function of temperature, including all energy levels in \citet{2019MNRAS.483..654S} and used by
\citet{2021ApJ...907...12S} in their study of high redshift quasars.
There is no cut-off criterion and we assume $W_n=1$.
The highest temperature plotted is that at which Fe is collisionally ionized to form Fe~III. 
To limit the temperature range, we used the condition 
Fe~II~$/$~Fe~$\leq~0.01$ and the ionization distribution 
represented in \citet{2013ApJ...768...82N}.
According to Figure~\ref{figs: U(FeII)}, the partition function for Fe~II is not severely divergent at the highest temperatures shown.
Its evaluation is not as simple as $1$ \& $2$ electron systems because it depends on the temperature across the entire range. 

\begin{figure}
\centering
\includegraphics[scale=0.8]{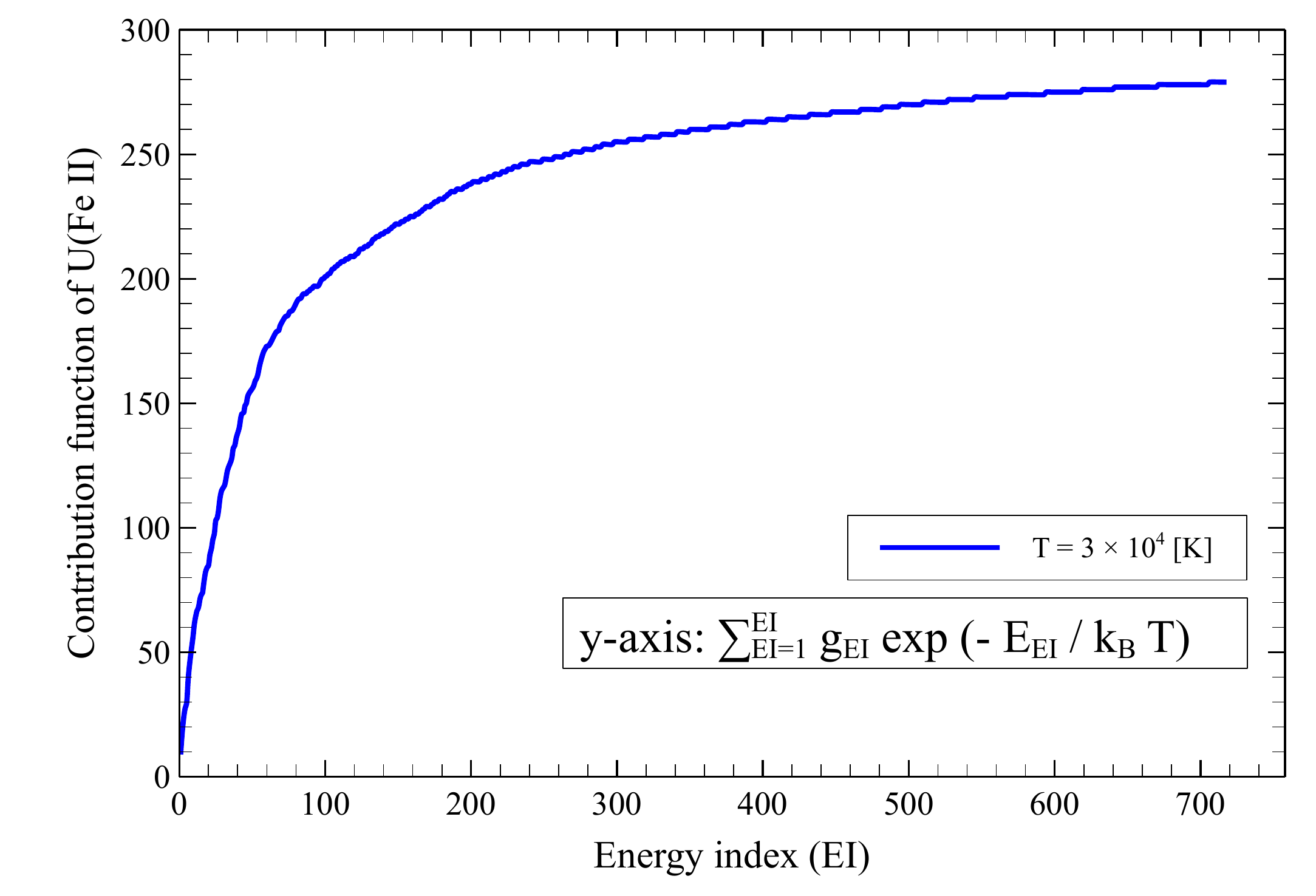}
\caption
{$U$ for Fe~II as a function of the number of levels included in the summation.
It is evaluated for $U(\mathrm{Fe~II})$ at $T=3\e{4}$~K.
The number of levels included in  Equation 
\ref{eq: Revised partition function_sharp_cutoff} is shown on
the x-axis.
Continuum lowering at high densities decreases the number
of levels.} 
\label{figs: Cont. FeII}
\end{figure}

Figure~\ref{figs: Cont. FeII} shows how $U(\mathrm{Fe~II})$ 
depends on the number of levels in Equation 
\ref{eq: Revised partition function_sharp_cutoff}.
This sum includes all states up to the energy index $EI$
shown as the x-axis. 
$U(\mathrm{Fe~II})$ is evaluated at $T = 3 \e{4}$ K since this
is the most extreme case with the largest $U$.
Continuum lowering  limits $EI$ so Figure \ref{figs: Cont. FeII}
illustrates how $U$ depends on density, assuming a density and
temperature can be converted into $EI$.

The partition function sum has not fully converged since it is still increasing at the highest energy index.
This suggests that highly excited and auto-ionizing energy levels
present in the real atom but not in the
\citet{2019MNRAS.483..654S} calculation might increase $U$ farther.
Figure \ref{figs: EnergyLevels} shows that many data sets have
a gap between the highest listed level and the continuum above.
There are actually an infinite number of very highly-excited levels 
in this gap. 
Sophisticated models of $U$, for instance \citet{2002AcA....52..195H,2001AcA....51..347H},
use atomic theory to account for these high levels and fill
in this gap.

Extensive calculations of $U$ for a wide variety of ions
have been done. These often combine 
accurate experimental energy levels 
and use atomic theory to account for missing levels.
Some examples include 
\citet{1998Ap&SS.259..173M} for H to Na,
\citet{1984A&AS...57...43H} for Fe and
\citet{2001AcA....51..347H} for Ni.
Many papers present the equivalent of Figure
\ref{figs: Cont. FeII}
with the relationship between continuum truncation 
and density/temperature left as a separate problem.
Table \ref{tab:HiZ-U} lists examples.
\begin{table}
 \centering
 \caption{\label{tab:HiZ-U}Partition functions 
 for complex ions}
 \null\smallskip
 \footnotesize
 \renewcommand\arraystretch{0.65}
 \begin{tabular}{ c c }
 \hline\hline
 Elements & reference \\
 \hline
 H to Na & 1 \\
 Ti, V & 2 \\
 Cr & 3 \\
Fe & 4 \\
Mn, Co & 5 \\
Ni & 6\\
 \hline
 \end{tabular}
\tablecomments{References; 1 \citet{1998Ap&SS.259..173M} ,
2 \citet{1988A&AS...75...47H},
3 \citet{1986A&AS...64..495H} ,
4 \citet{1984A&AS...57...43H} , 
5 \citet{1989A&AS...81..303H},
6 \citet{2001AcA....51..347H}}
 \end{table}

There is no uncertainty in evaluating $U$ as a function of
the energy index in Figure~\ref{figs: Cont. FeII}. 
To find the highest energy index to include in the $U\mathrm{(Fe~II)}$ sum, we need a truncation theory to map this index to the density and temperature effects in the dense plasma. 
Therefore, the next step is to evaluate $U\mathrm{(Fe~II)}$ for realistic densities,
which is equivalent to understanding the energy index in Figure~\ref{figs: Cont. FeII} where $W_n$ becomes small. 

The challenge is to convert a particular density and temperature into the occupation probability or $\mathrm{n_{max}}$. 
Here, we consider the case of Fe~II at $T = 3\e{4}$~K and 
a continuum lowering $4.6~\mathrm{eV}$.
$\Delta \chi = 4.6~\mathrm{eV}$ below the ionization limit corresponds to energy index of $EI = 400$ in Figure \ref{figs: Cont. FeII}.
This is a maximum quantum number of $\mathrm{n_{max}} \approx 49$ in Equation~\ref{eq: Revised partition function}
using the hydrogenic approximation.
Equation \ref{eq: delta_x Mihalas}
gives a required density of
$N_{\mathrm{e}} \approx 1.74 \e{20}\ \mathrm{cm^{-3}}$
to lower the continuum by this amount. 
One should be careful in applying the hydrogenic criterion 
to many-electron ions since the energy levels of the hydrogen-like species 
are explicitly related to the principal quantum number 
while this is not the case for many-electron ions 
unless we assume the hydrogenic energy level structure 
for these ions as well.

We discussed in Section \ref{Inter-particle interactions with Sharp cut-off},
the 
continuum-lowering theories presented by
\citet{1998Ap&SS.259..173M, 2000ApJ...544..581B, 2012fapc.book.....C}
give different equations with different density dependencies. 
Equations ~$3$,~$4$,~and~$5$  in \citet{2000ApJ...544..581B} give the densities of
$N_{\mathrm{e}} = 8.55 \e{18}~\mathrm{cm^{-3}}$,
$N_{\mathrm{e}} = 6.09 \e{29}~\mathrm{cm^{-3}}$,
$N_{\mathrm{e}} = 2.44 \e{15}~\mathrm{cm^{-3}}$, respectively.
More recently, \citet{2020arXiv201110603K} give a similar review of these equations.
\citet{1998Ap&SS.259..173M}
gives a density of
$N_{\mathrm{e}} = 7.09 \e{19}~\mathrm{cm^{-3}}$.
Equation~$8.4$ presented in \citet{2012fapc.book.....C} gives the density of
$N_{\mathrm{e}} =5.2 \e{13}~\mathrm{cm^{-3}}$.
And finally, Equation~\ref{eq: maximum n} for $\mathrm{n_{max}}=49$ gives the density of
$N_{\mathrm{e}} =5.12 \e{11}~\mathrm{cm^{-3}}$.
These theories are uncertain for the case of Fe~II, as shown by the large discrepancies
in the required densities, and this, in turn, is due
to fundamental questions
about plasma effects \citep{2005pps..book.....G}.

Our case study of Fe~II shows that while it is simple to
specify the partition function as a number of
levels, converting a particular density and temperature into
a limit on the number of levels is uncertain.
Extensive studies have been done showing
$U$ for most astrophysically abundant elements
as a function of temperature and the highest level, or the
degree of continuum lowering.
We highlight two representative studies that evaluate $U$ versus continuum lowering with no density dependency.
\citet{2002AcA....52..195H} studied Fe~IV, (their figure~$1$ and table~$1$)
for the temperature range where it is abundant.
They include the higher energy levels missing from experiment but
predicted by quantum mechanics,
including auto-ionizing energy levels.
They show how $U(\mathrm{Fe~IV})$ varies with continuum-lowering energy and temperature. The same evaluation for Ni~IV is shown in figure~$1$ presented in~\citet{2001AcA....51..347H}.

\citet{2013PhPl...20c2108D} shows another example using the pressure-lowering criterion.
They evaluate the partition function for carbon, C~I - C~IV, using the three so-called lumped levels of the ground state, low-lying excited states called \nth{1} excited states, and the rest called \nth{2} excited states.
The result is shown in figure~$1$ of the paper.
Like the partition function of Fe~IV, the partition function is evaluated versus temperature and continuum lowering.
These authors do not attempt to convert a temperature 
or density into the explicit continuum lowering 
that will occur.
That last step is the essential difficulty with only 
discrepant theories available to the worker.

These issues are also present in laboratory plasmas like
Tokamaks \citep{2016mmcr.book.....R}.  
Here, spectra are used to probe conditions in the plasma,
so getting the ``right answer'' is important.
The dense-plasma community organizes a series of ``NLTE'' workshops, which gather developers of spectral simulation codes to compare results and discuss physical methods.
We have participated in two of these workshops, summarized by
\citet{2013HEDP....9..645C} and
\citet{2017HEDP...23...38P}, and presented predictions of the spectral
synthesis code Cloudy \citep{2017RMxAA..53..385F}.
\citet{2020arXiv201110603K} include predictions of XSTAR \citep{2000ApJ...544..581B} in another workshop.
The plasma codes are not in good agreement at high densities
due to various treatments of collisional effects upon
highly excited states.
These and other dense-plasma challenges are reviewed by \citet{2016mmcr.book.....R}.
\newpage
\section{The molecular partition function}
\label{A short note about the molecular partition function}

We do not discuss the molecular partition function,
but we could extend our results based on the energy structure
to examine this behavior.
Molecules have a very different energy-level structure, making their properties very different from what we have described for simple or complex ions.
The essential difference is that molecules do not have an infinite number of rotational-vibrational levels. The rotational energy-level spacing increases as the energy increases, so there are only a few highly excited states.
This property of the energy structure of molecules
is shown in figures~$2$~and~$3$ presented in \citet{2005ApJ...624..794S}
for $\mathrm{H}_2$, which can be compared with
Figure \ref{figs: EnergyLevels}. 
The divergence and truncation
of the molecular partition function is not a central concern
for these low levels.
For further discussions about the molecular partition function,
the reviews by \citet{1984ApJS...56..193S, 2015PASP..127..266M} are useful.
Polynomial expansions for diatomic molecules of astrophysical
interest are given  by
\citet{1984ApJS...56..193S}.
 
\section{Discussion, summary, and conclusion}
\label{sec: Discussion, summary, and conclusion}

This tutorial offers an introduction to numerical calculations
of the partition function $U$.  
Accurate values are needed to predict the populations or
ionization of a gas in thermodynamic equilibrium or 
local thermodynamic equilibrium.  
Although texts do cover the definition of $U$, and some
give it as the statistical weight of the ground state, 
we know of none which provide practical advice 
on obtaining numerical values for $U$ for a particular density
and temperature.

Our major points are:
\begin{itemize}
  \item The partition is infinite for 
  any atom with an infinite number of levels.
  This is referred to as the divergence of the partition function.
  Although the principal quantum number $n$ can extend
   to infinity in quantum mechanics, the radius of
  the orbit also goes to infinity.
  Large orbits are not possible since
  an atom would overlap with its neighbors.
  There must be the largest possible orbit that would
  be smaller for a denser gas.  
  This introduces the concept of a density-dependent
  cutoff or truncation to the number of levels included 
  in Equation \ref{eq: Partition function}.
  This is often referred to as continuum lowering at
  high densities.

\item Calculating $U$ is relatively simple for
  one and two electron atoms like hydrogen or helium,
  because their orbits have the hydrogenic spacing 
  shown in Figure \ref{figs: EnergyLevels}.
  The larger orbits which cause $U$ to diverge are
  very close to the continuum.  
  Temperatures high enough to populate the highest levels
  would also ionize the atom so there is an
  upper limit to the temperature where the atom
  is abundant and $U$ matters.
  By combining the Saha and Boltzmann equations we show
  that for most conditions $U$ is the statistical
  weight of the ground state.  It increases to modest
  values, generally $U<10$, for a narrow range of
  density and temperature.
  
\item The  energy-level structure of many-electron systems 
  like Fe~II, also
  shown in Figure \ref{figs: EnergyLevels}, is far more 
  complex.
  We need a theory to specify how many levels to include
  in Equation \ref{eq: Partition function}.
  Figure \ref{figs: U(FeII)} shows $U$ for Fe~II over a
  range of temperature while Figure \ref{figs: Cont. FeII}
  shows the effects of varying the number
  of levels, which is equivalent to varying the density
  since continuum lowering is more severe at high densities.  
  
\item Two approaches are taken to truncate the 
  number of levels that contribute to $U$.  The
  hard-sphere model 
  establishes a largest level that can
  exist in the space available between atoms and ions.  This sets
  an upper limit to the sum in Equation \ref{eq: Partition function} as given in Equation \ref{eq: Revised partition function_sharp_cutoff}.
  The occupation probability method assigns a likelihood 
  $W_n$ that
  a level is occupied. The sum becomes Equation \ref{eq: Revised partition function}.
  Both approaches are used.
  
\item The stellar astrophysics community has adopted the
  approach of \citet{1988ApJ...331..794H},
  with over six hundred citations listed in the ADS.
  They describe the dissolution of the higher
  energy levels with the Stark effect.
  Other theories are used in other communities
  such as accretion flows near black holes \citep{2020arXiv201110603K}
  or a
  laboratory plasma \citep{2016mmcr.book.....R}.
  The various theories differ by large amounts
  giving densities that can range over more than one 
  dex for a given truncation.
  This shows the difficulty in treating the effects of
  the sea of free electrons that surround an atom in
  a dense gas.
  This is a long-standing and vexing problem that spans
  many parts of physics.
  
\end{itemize}

\acknowledgments

GJF acknowledges support by NSF (1816537, 1910687), NASA (ATP 17-ATP17-0141, 19-ATP19-0188), and STScI (HST-AR- 15018).

\addcontentsline{toc}{section}{References}
\bibliography{LocalBibliography}

\begin{thebibliography}{}
\expandafter\ifx\csname natexlab\endcsname\relax\def\natexlab#1{#1}\fi
\providecommand{\url}[1]{\href{#1}{#1}}
\providecommand{\dodoi}[1]{doi:~\href{http://doi.org/#1}{\nolinkurl{#1}}}
\providecommand{\doeprint}[1]{\href{http://ascl.net/#1}{\nolinkurl{http://ascl.net/#1}}}
\providecommand{\doarXiv}[1]{\href{https://arxiv.org/abs/#1}{\nolinkurl{https://arxiv.org/abs/#1}}}

\bibitem[{{Allen}(1973)}]{1973asqu.book.....A}
{Allen}, C.~W. 1973, {Astrophysical quantities}

\bibitem[{{Bautista} \& {Kallman}(2000)}]{2000ApJ...544..581B}
{Bautista}, M.~A., \& {Kallman}, T.~R. 2000, \apj, 544, 581,
  \dodoi{10.1086/317206}

\bibitem[{Blundell \& Blundell(2010)}]{blundell2010concepts}
Blundell, S., \& Blundell, K. 2010, Concepts in Thermal Physics (OUP Oxford).
\newblock \url{https://books.google.com/books?id=T0luBAAAQBAJ}

\bibitem[{{Bradt}(2014)}]{2014aspr.book.....B}
{Bradt}, H. 2014, {Astrophysics Processes}

\bibitem[{{Capitelli} {et~al.}(2012){Capitelli}, {Colonna}, \&
  {D'Angola}}]{2012fapc.book.....C}
{Capitelli}, M., {Colonna}, G., \& {D'Angola}, A. 2012, {Fundamental Aspects of
  Plasma Chemical Physics}, Vol.~66, \dodoi{10.1007/978-1-4419-8182-0}

\bibitem[{{Carroll} \& {Ostlie}(2006)}]{1996ima..book.....C}
{Carroll}, B.~W., \& {Ostlie}, D.~A. 2006, {An Introduction to Modern
  Astrophysics}

\bibitem[{{Chung} {et~al.}(2013){Chung}, {Bowen}, {Fontes}, {Hansen}, \&
  {Ralchenko}}]{2013HEDP....9..645C}
{Chung}, H.~K., {Bowen}, C., {Fontes}, C.~J., {Hansen}, S.~B., \& {Ralchenko},
  Y. 2013, High Energy Density Physics, 9, 645,
  \dodoi{10.1016/j.hedp.2013.06.001}

\bibitem[{{Cox}(2000)}]{2000asqu.book.....C}
{Cox}, A.~N. 2000, {Allen's astrophysical quantities}

\bibitem[{{D'Ammando} {et~al.}(2013){D'Ammando}, {Colonna}, \&
  {Capitelli}}]{2013PhPl...20c2108D}
{D'Ammando}, G., {Colonna}, G., \& {Capitelli}, M. 2013, Physics of Plasmas,
  20, 032108, \dodoi{10.1063/1.4794286}

\bibitem[{{de Galan} {et~al.}(1968){de Galan}, {Smith}, \&
  {Winefordner}}]{1968AcSpe..23..521D}
{de Galan}, L., {Smith}, R., \& {Winefordner}, J.~D. 1968, Spectrochimica Acta,
  23, 521, \dodoi{10.1016/0584-8547(68)80032-1}

\bibitem[{{de Jager} \& {Neven}(1960)}]{1960BAN....15...55D}
{de Jager}, C., \& {Neven}, L. 1960, \bain, 15, 55

\bibitem[{{Ferland} {et~al.}(2017){Ferland}, {Chatzikos}, {Guzm{\'a}n},
  {Lykins}, {van Hoof}, {Williams}, {Abel}, {Badnell}, {Keenan}, {Porter}, \&
  {Stancil}}]{2017RMxAA..53..385F}
{Ferland}, G.~J., {Chatzikos}, M., {Guzm{\'a}n}, F., {et~al.} 2017, Revista
  Mexicana de Astronomia y Astrofisica, 53, 385.
\newblock \doarXiv{1705.10877}

\bibitem[{{Griem}(2005)}]{2005pps..book.....G}
{Griem}, H.~R. 2005, {Principles of Plasma Spectroscopy}

\bibitem[{{Hahn}(1997)}]{1997PhLA..231...82H}
{Hahn}, Y. 1997, Physics Letters A, 231, 82,
  \dodoi{10.1016/S0375-9601(97)00287-9}

\bibitem[{{Halenka}(1988)}]{1988A&AS...75...47H}
{Halenka}, J. 1988, \aaps, 75, 47

\bibitem[{{Halenka}(1989)}]{1989A&AS...81..303H}
---. 1989, \aaps, 81, 303

\bibitem[{{Halenka} \& {Grabowski}(1984)}]{1984A&AS...57...43H}
{Halenka}, J., \& {Grabowski}, B. 1984, \aaps, 57, 43

\bibitem[{{Halenka} \& {Grabowski}(1986)}]{1986A&AS...64..495H}
---. 1986, \aaps, 64, 495

\bibitem[{{Halenka} \& {Madej}(2002)}]{2002AcA....52..195H}
{Halenka}, J., \& {Madej}, J. 2002, \actaa, 52, 195.
\newblock \doarXiv{astro-ph/0204384}

\bibitem[{{Halenka} {et~al.}(2001){Halenka}, {Madej}, {Langer}, \&
  {Mamok}}]{2001AcA....51..347H}
{Halenka}, J., {Madej}, J., {Langer}, K., \& {Mamok}, A. 2001, \actaa, 51, 347.
\newblock \doarXiv{astro-ph/0201238}

\bibitem[{{Hubeny} \& {Mihalas}(2014)}]{2014tsa..book.....H}
{Hubeny}, I., \& {Mihalas}, D. 2014, {Theory of Stellar Atmospheres}

\bibitem[{{Hummer}(1986)}]{1986JQSRT..36....1H}
{Hummer}, D.~G. 1986, \jqsrt, 36, 1, \dodoi{10.1016/0022-4073(86)90011-7}

\bibitem[{{Hummer} \& {Mihalas}(1988)}]{1988ApJ...331..794H}
{Hummer}, D.~G., \& {Mihalas}, D. 1988, \apj, 331, 794, \dodoi{10.1086/166600}

\bibitem[{{Inglis} \& {Teller}(1939)}]{1939ApJ....90..439I}
{Inglis}, D.~R., \& {Teller}, E. 1939, \apj, 90, 439, \dodoi{10.1086/144118}

\bibitem[{{Kallman} {et~al.}(2020){Kallman}, {Bautista}, {Deprince}, {Garcia},
  {Mendoza}, {Ogorzalek}, {Palmeri}, \& {Quinet}}]{2020arXiv201110603K}
{Kallman}, T., {Bautista}, M., {Deprince}, J., {et~al.} 2020, arXiv e-prints,
  arXiv:2011.10603.
\newblock \doarXiv{2011.10603}

\bibitem[{{Kramida} {et~al.}(2014){Kramida}, {Ralchenko}, \&
  {Reader}}]{2014APS..DMP.D1047K}
{Kramida}, A., {Ralchenko}, Y., \& {Reader}, J. 2014, in APS Division of
  Atomic, Molecular and Optical Physics Meeting Abstracts, APS Meeting
  Abstracts, D1.047

\bibitem[{{Mangum} \& {Shirley}(2015)}]{2015PASP..127..266M}
{Mangum}, J.~G., \& {Shirley}, Y.~L. 2015, \pasp, 127, 266,
  \dodoi{10.1086/680323}

\bibitem[{{Mihalas}(1978)}]{1978stat.book.....M}
{Mihalas}, D. 1978, {Stellar atmospheres}

\bibitem[{{Milone} \& {Merlo}(1998)}]{1998Ap&SS.259..173M}
{Milone}, L.~A., \& {Merlo}, D.~C. 1998, \apss, 259, 173,
  \dodoi{10.1023/A:1001508021614}

\bibitem[{{Nikoli{\'c}} {et~al.}(2013){Nikoli{\'c}}, {Gorczyca}, {Korista},
  {Ferland }, \& {Badnell}}]{2013ApJ...768...82N}
{Nikoli{\'c}}, D., {Gorczyca}, T.~W., {Korista}, K.~T., {Ferland }, G.~J., \&
  {Badnell}, N.~R. 2013, \apj, 768, 82, \dodoi{10.1088/0004-637X/768/1/82}

\bibitem[{{Novotny}(1973)}]{1973itsa.book.....N}
{Novotny}, E. 1973, {Introduction to stellar atmospheres and interiors}

\bibitem[{{Osterbrock} \& {Ferland}(2006)}]{2006agna.book.....O}
{Osterbrock}, D.~E., \& {Ferland}, G.~J. 2006, {Astrophysics of gaseous nebulae
  and active galactic nuclei}

\bibitem[{{Piron} {et~al.}(2017){Piron}, {Gilleron}, {Aglitskiy}, {Chung},
  {Fontes}, {Hansen}, {Marchuk}, {Scott}, {Stambulchik}, \&
  {Ralchenko}}]{2017HEDP...23...38P}
{Piron}, R., {Gilleron}, F., {Aglitskiy}, Y., {et~al.} 2017, High Energy
  Density Physics, 23, 38, \dodoi{10.1016/j.hedp.2017.02.009}

\bibitem[{{Ralchenko}(2016)}]{2016mmcr.book.....R}
{Ralchenko}, Y. 2016, {Modern Methods in Collisional-Radiative Modeling of
  Plasmas}

\bibitem[{{Sarkar} {et~al.}(2021){Sarkar}, {Ferland}, {Chatzikos},
  {Guzm{\'a}n}, {van Hoof}, {Smyth}, {Ramsbottom}, {Keenan}, \&
  {Ballance}}]{2021ApJ...907...12S}
{Sarkar}, A., {Ferland}, G.~J., {Chatzikos}, M., {et~al.} 2021, \apj, 907, 12,
  \dodoi{10.3847/1538-4357/abcaa6}

\bibitem[{{Sauval} \& {Tatum}(1984)}]{1984ApJS...56..193S}
{Sauval}, A.~J., \& {Tatum}, J.~B. 1984, \apjs, 56, 193, \dodoi{10.1086/190980}

\bibitem[{{Shaw} {et~al.}(2005){Shaw}, {Ferland}, {Abel}, {Stancil}, \& {van
  Hoof}}]{2005ApJ...624..794S}
{Shaw}, G., {Ferland}, G.~J., {Abel}, N.~P., {Stancil}, P.~C., \& {van Hoof},
  P.~A.~M. 2005, \apj, 624, 794, \dodoi{10.1086/429215}

\bibitem[{{Smyth} {et~al.}(2019){Smyth}, {Ramsbottom}, {Keenan}, {Ferland }, \&
  {Ballance}}]{2019MNRAS.483..654S}
{Smyth}, R.~T., {Ramsbottom}, C.~A., {Keenan}, F.~P., {Ferland }, G.~J., \&
  {Ballance}, C.~P. 2019, \mnras, 483, 654, \dodoi{10.1093/mnras/sty3198}

\bibitem[{{Swihart}(1968)}]{1968asa..book.....S}
{Swihart}, T.~L. 1968, {Astrophysics and stellar astronomy}

\bibitem[{{Uns{\"o}ld}(1948)}]{1948ZA.....24..355U}
{Uns{\"o}ld}, A. 1948, \zap, 24, 355

\bibitem[{{Vera Rueda} \& {Rohrmann}(2020)}]{2020A&A...635A.180V}
{Vera Rueda}, M., \& {Rohrmann}, R.~D. 2020, \aap, 635, A180,
  \dodoi{10.1051/0004-6361/201937413}

\end{thebibliography}

\end{document}